\documentclass[%
 aip,
 jmp,%
 amsmath,amssymb,
 reprint,%
]{revtex4-1}

\usepackage{graphicx}
\usepackage{dcolumn}
\usepackage{bm}
\newcommand {\al}   {\alpha}          \newcommand {\bt}  {\beta}
\newcommand {\g }   {\gamma}          \newcommand {\G }  {\Gamma}
\newcommand {\dl}   {\delta}          
           
          \renewcommand {\k }  {\kappa}
         
\newcommand {\n }   {\nu}             
\renewcommand {\r }   {\rho}            
            \newcommand {\ph}  {\phi}
\newcommand {\h }  {\chi}
\newcommand {\pf}   {\psi}            \newcommand {\om}  {\omega}
         \newcommand {\nb}  {\nabla}
\newcommand {\pl}   {\partial}        \newcommand {\Om}  {\Omega}

  \newcommand {\be}{\begin{equation}\label}   \newcommand {\ee}{\end{equation}}
\newcommand {\bear}{\begin{eqnarray}\label}   \newcommand {\eear}{\end{eqnarray}}
\newcommand {\bean}{\begin{eqnarray*}}   \newcommand {\eean}{\end{eqnarray*}}
\newcommand {\ba}{\begin{array}}   \newcommand {\ea}{\end{array}}
       \newcommand{\lb}{\label}
 
        \def\({\left(} \def\){\right)}
      \def\l[{\left[} \def\r]{\right]}

\def\bfh{\h\!\!\!\!\hspace{0.125pt}\h \!\!\!\!\h}

\def\bfom{\om\!\!\!\!\hspace{0.2pt}\om \!\!\!\!\hspace{0.3pt}\om}

\begin{document}

\preprint{AIP/123-QED}

\title[]{FOUNDATION OF THE MECHANICS OF ORIENTED POINT}

\author{Trukhanova Mariya and Shipov Gennady}
 \altaffiliation[ ]{M.V.Lomonosov Moscow State University, Faculty of Physics, Leninskie Gory,  Moscow, Russia}
\email{mar-tiv@yandex.ru}

\date{02.01.2013}

\begin{abstract}
The mechanics of an oriented  point (point with "spin") based on 3D and 4D Frenet equations is considered. In such mechanics there is an opportunity to describe formally any physical trajectory of a particle with own rotation.
We use  anholonomic rotational coordinates (Euler angles) as elements of internal space  of the mechanics  which generate a rotational relativity. The  groups of transformations of the mechanics of an oriented  point form Poincare's group with semidirect product of translations and rotations, so translational and rotational momentums appear dependent from each other. Connection of the curve torsion  with Ricci rotational coefficients is shown and
rotational metric is entered. Equivalence between equations of motion 4D oriented point and geodesic equations of absolute parallelism geometry is established. The space of events an arbitrary   accelerated 4D frame of reference, which has 10 degrees of freedom, is described by Cartan structure equations of absolute parallelism geometry $A_4(6)$. It represent 10D coordinate space in which 4 translational  coordinates $x_0=ct,~x_1=x,~x_2=y,x_3=z$  describe motion of the origin $O$ 4D oriented  point and  6 angular coordinates $\varphi_1=\varphi,~\varphi_2=\psi,~\varphi_3=\theta,~\varphi_4=\vartheta_x,~\varphi_5=\vartheta_y,~\varphi_6=\vartheta_z$ describe change of its orientation. The structural equations of absolute parallelism  geometry $A_4(6)$,
represent an extended set of Einstein-Yang-Mills equations with the gauge
translations group $T_4$ defined on the base $x^i$  and with the gauge rotational group $O(1.3)$, defined in the
fibre $e^i_{~a}$.  The sources in these equations are defined through the torsion (torsion field)
of $A_4(6)$ geometry. The received system of the equations represents generalization vacuum Einstein's equations on a case when sources have  geometrical nature. On the basis of the Vaidya-like solution of the Einstein-Yang-Mills equations correspondence with the Einstein's equations is established. \end{abstract}

\maketitle
\section{Introduction}

   Einstein proposed the following approach to the construction of the fun\-da\-men\-tal
theory of things in the microworld. On the left-hand side of his famous
equations

\begin{equation}
                         \label{1}
R_{jm}-\frac{1}{2}g_{jm}R=\frac{8\pi G}{c^{4}}T_{jm}
\end{equation}
one finds a purely geometrical quantity  (the Einstein tensor
$G_{jm}=R_{jm}- \frac{1}{2}g_{jm}R$), and on the right-hand side, the energy-momentum
tensor $T_{jm}$ for mater, which was, so to speak, introduced "manually."
In Einstein's picture, mater appears thus against the background of a curved
space-time as an entity independent of  space-time.

Einstein was not satisfied with the phenomenological representation of
$T_{jm}$, since: "The right-hand side includes all that cannot be
described so far in the unified field theory. Of course, not for a fleeting
moment I have had any doubt that such a formulation is just a temporary answer,
undertaken to give to general relativity some closed expression. This
formulation has been in essence nothing more that the theory of the gravitation
field, which has been separated in a somewhat artificial manner from the
unified field of a yet unknown nature  \cite{1}."

A way to remove arbitrariness in the selection of the energy-momentum tensor
was seen by Einstein in the geometrization of the energy-momentum tensor of
matter on the right-hand side of Einstein's equations (\ref{1}).
Einstein believed that the geometrization of the energy-momentum
tensor of matter should result in the geometrization of the matter field that
make it up. For Einstein the geometrization of matter fields implied the
construction of a fun\-da\-men\-tal theory of phenomena in the microworld that is
in conformity with relativity principle.  30 years he tried
constructing a "reasonable general relativistic theory," and within its
framework a "more advanced quantum theory \cite{2}."

In the present work we offer to geometrize the right part of the equations (\ref{1}), using instead of a material point of general relativistic   Einstein's mechanics
more the general object - an oriented  material point. We shall understand any  accelerated reference frame,  formed by unit orthogonal vectors as an oriented  material point. In 3D translational coordinates space the oriented point has 6 degrees of freedom, in 4D translational coordinate space - ten. Generalization of the  Einstein's  theory offered by us allows analytically to describe Descartes's approved idea, that any real motion is a rotation.

\vspace{4mm} {\raggedright \section{3D oriented point and
generalization of the equations of Newton's  mechanic }} \vspace{4mm}

\begin{figure}[htbp]
\centering\includegraphics[width=100mm,height=60mm]{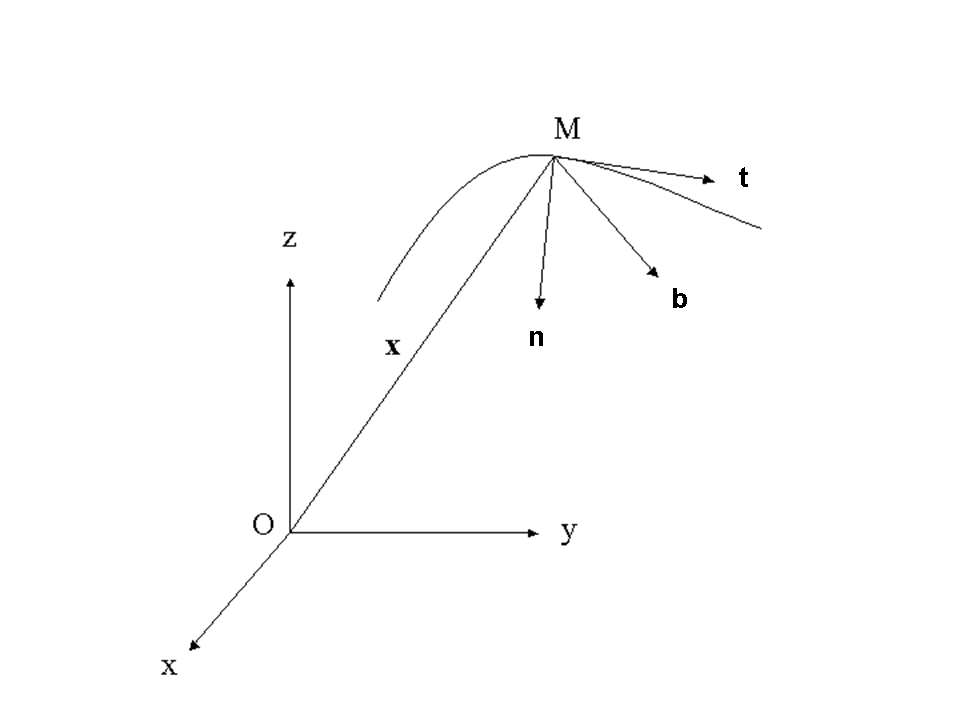}
\caption{\label{r1}Trajectory of  the 3D oriented point}
\end{figure}

In 1847 French mathematician Jean F. Frenet  in his thesis
   has
written  equations, describing motion of oriented
point in the 3D space along  an arbitrary curve $ {\bf x}= {\bf x}(s)$, where $s$ \---
the arc length. Equations are written for the three  orthogonal unit
vectors  ${ \bf t},~  {\bf n}$ and ${\bf b}$ with orthogonality conditions ${\bf t}^2={\bf n}^2={\bf b}^2=1,~~{\bf t}{\bf n}= {\bf n}{\bf b}={\bf b}{\bf t}=0.$ The tangent unit vector ${\bf t}$ is choused as tangent to the curve at point M (fig.\ref{r1}), pointing the direction of motion. The normal unit vector ${\bf n}$ is the derivative of  ${\bf t}$ with respect to the arc length parameter of the curve, divided by its length and the binormal unit vector ${\bf b}$ is defined as the cross product ${\bf b}={\bf t}\times{\bf n}$.

For these vectors Frenet's equations look like  \cite{3}
\be{2}
\frac{d {\bf x}}{ds}={\bf t}. \ee
\be{3} \frac{d {\bf t}}{ds}=\k (s){ \bf n}, \ee
\be{4} \frac{d{\bf n}}{ds}=-\k (s){\bf t}+ \h (s){\bf b}, \ee
\be{5} \frac{d {\bf b}}{ds}=- \h (s){\bf n}, \ee
where $\k (s)$ - curvature of the curve and  $\h(s)$ - torsion of the curve.
Frenet was the first who has shown that arbitrary curve in 3D flat space is
 determined by two scalar parameters - curvature $\k (s)$ and
torsion $\h(s)$.

 Differentiating  equations (\ref{2})  on $s$ and using (\ref{3})-(\ref{5}) and
orthogonality conditions, we shall get the equations

\be{6}\frac{d^2 {\bf x}}{ds^2}= \k (s){\bf n}~, \ee
\be{7}\frac{d^3{\bf x}}{ds^3}= \frac{d\k (s)}{ds}{\bf n}-
\k^2 (s){\bf t} + \k (s)\h (s){\bf b}~, \ee
describing motion of the triad origan $O$ (motion of point M).

For comparison of the equations (\ref{6}) è (\ref{7}) with the equations of Newton's mechanics, it is convenient to pass in them to time parameter  $t$
\be{8}\frac{d^2 {\bf x}}{dt^2}=a{\bf t} + \k v^2{\bf n}~, \ee
\be{9}\frac{d^3 {\bf x}}{dt^3}=
(\frac{da}{dt}-\k^2 v^3){\bf t} + (3av\kappa+v^2\frac{d\kappa}{dt}){\bf n}+\k \h v^3{ \bf b}~,
\ee
 where $v=ds/dt$ \--- absolute velocity  and $a=dv/dt$\--- tangent acceleration.

  Multiplying these equations on mass $m$, we shall receive the translational equations of motion of an oriented point with the law of transformation infinitesimal  vector $dx_\alpha$
$$dx_{\alpha'}=\frac{\partial x_{\alpha'}}{\partial x_\alpha}dx_\alpha,~~~\alpha=1,2,3,$$
  where matrixes $\partial x_{\alpha'}/\partial x_\alpha$ form the group 3D translations $t(3)$. The equations (\ref{8}) are similar to the equations of  Newton mechanics, but have  geometrical nature. A choice of curvature  $\kappa$ and parameter  $s$ it is possible to describe formally any physical trajectory of a particle in 3D space, moving under action of force ${\bf F}=m (a{\bf t} + \k v^2{\bf n}).$  The equations (\ref{9}) have no analogues in the Newton mechanic   as contain the third derivative of coordinate on time. The system of the equations (\ref{8})è (\ref{9}) describes the motion of the origin of an oriented material point
 taking into account of its spin and, certainly, generalizes the equations of Newton's mechanics.

\vspace{4mm} {\raggedright \section{Internal  space of the anholonomic rotational coordinates}} \vspace{4mm}

\begin{figure}[h]
\centering\includegraphics[width=120mm,height=80mm]{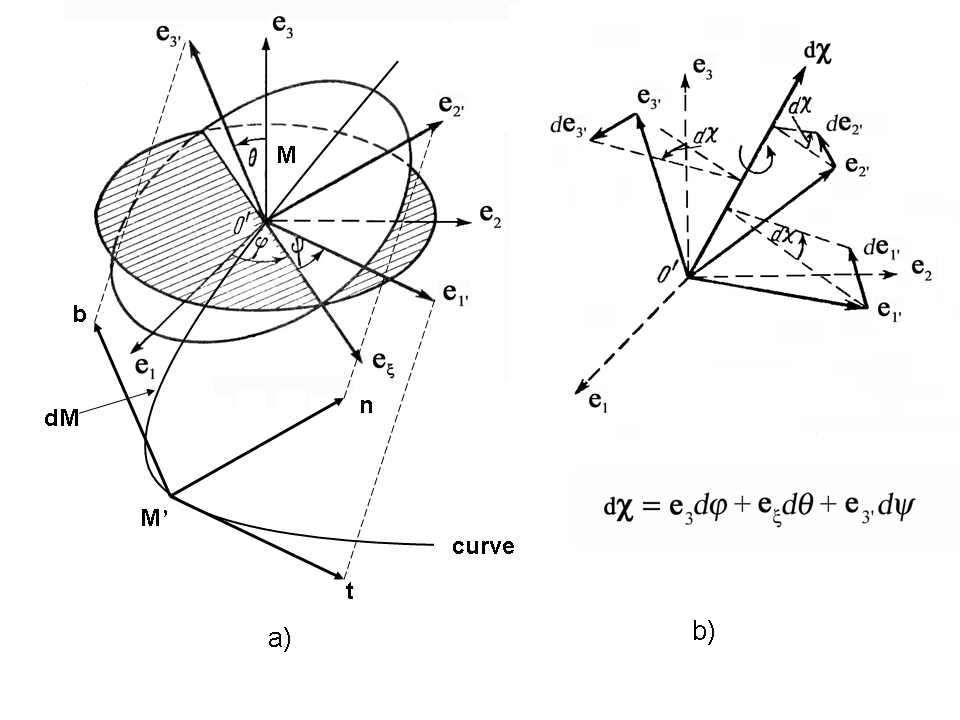}
\caption{\label{r2} a)~Changing of the orientation of an oriented point at displacement  of point M on dM;~b)~according to Euler's theorem  an infinitesimal rotations around
the three axes it is possible to replace  by one rotation with the an infinitesimal angle}
\end{figure}

 During infinitesimal displacement of point M along the curve
the triad of Frenet's vectors   simultaneously change their orientation in
space. For description of the change it is convenient to introduce
anholonomic angular coordinates
$$\varphi=\angle({\bf e}_1~{\bf e}_\xi),~~~\psi=\angle({\bf e}_\xi~{\bf e}_{1'}),~~~\theta\angle({\bf e}_3~{\bf e}_{3'}),$$
$$(0\leq\varphi\leq 2\pi,~~~~0\leq\psi\leq 2\pi,~~~~0\leq \theta\leq \pi,)$$
 \--- Euler angles (see  (fig.\ref{r2}a)).

Let's assume, that with a point  $M$ of a curve the triad  with motionless unit vectors ${\bf e}_1, ~{\bf e}_2,~ {\texttt{}\bf e}_3$ is connected.
Let's designate components of motionless  Frenet triad  as
$${ \bf t}={\bf e}^{'}_1,~  {\bf n}={\bf e}^{'}_{2},~{\bf b}={\bf e}^{'}_{3}.$$

   At displacement of the origin $O$ of Frenet triad along a curve from a point $M $ in a point $M'$, there is a rotation vectors of Frenet triad  (fig. \ref {r2} a). Projecting the axes of a mobile triad
 ${\bf t},~ {\bf n},~ {\bf b} $, located in a point $M'$ on the motionless triad connected with a point $M $, we find
     \begin{widetext}
$$ {\bf t}={\bf e}^{'}_1={\bf e}_1(\cos\varphi\cos\psi-\sin\varphi\sin\psi\cos\theta)+{\bf e}_2(\sin\varphi\cos\psi+\cos\varphi\sin\psi\cos\theta)+{\bf e}_3\sin\psi\sin\theta,$$
$${\bf n}={\bf e}^{'}_{2}=-{\bf e}_1(\cos\varphi\sin\psi+\sin\varphi\cos\psi\cos\theta)-{\bf e}_2(\sin\varphi\sin\psi-\cos\varphi\cos\psi\cos\theta)+{\bf e}_3\cos\psi\sin\theta,$$
$${\bf b}={\bf e}^{'}_{3}={\bf e}_1\sin\varphi\sin\theta-{\bf e}_2\cos\varphi\sin\theta+{\bf e}_3\cos\theta.$$     \end{widetext}

 Expressing the components
of tangent vector ${\bf t}={\bf e'}{_1}=d{\bf x}/ds$ through angular variables, we have

\be{10} \frac{dx}{ds}=\cos{\varphi}~\cos{\pf} - \sin{\varphi}~ \sin{\pf}~ \cos\theta,\ee
\be{11} \frac{dy}{ds}=\sin{\varphi}~\cos{\pf} + \cos{\varphi}~\sin{\pf}~\cos\theta,\ee
\be{12} \frac{dz}{ds}=\sin{\pf}~\sin\theta.\ee

 Differentiating  the third components of vectors ${\bf t}$ and ${\bf n}$ and second component of the vector ${\bf b}$, we get "rotational equations of motion" as follows

\be{13} \frac{d\varphi}{ds}=\h\frac{\sin \pf}{\sin \theta}, \ee
\be{14} \frac{d\psi}{ds}= \kappa -\chi~\sin ~\psi ~ctg ~\theta, \ee
\be{15} \frac{d\ph}{ds}=\h~\cos \pf. \ee

According to the equations (\ref{13})-(\ref{15})  curvature $\kappa$ and torsion $\chi$  cause rotation of Frenet triad, therefore more correctly to name their first $\chi_1=\kappa$ and second $\chi_2=\chi$  torsion of a curve.

The system of the equations (\ref{10})-(\ref{15}) represents system Cauchy for six unknown functions $x,~y,~z,~\varphi,~\psi,~\theta$  nd supposes one and only one solution in the form of regular functions
$$x=x(s),~y=y(s),~z=z(s),~\varphi=\varphi(s),~\psi=\psi(s),~\theta=\theta(s)$$
satisfying to the equations (\ref{10})-(\ref{15})  and entry conditions
$$x=x_0,~y=y_0,~z=z_0,~\varphi=\varphi_0,~\psi=\psi_0,~\theta=\theta_0$$
 for $s=s_0.$  Entry conditions have simple geometrical sense. Initial coordinates $x=x_0,~y=y_0,~z=z_0$ define position of an original point  $M_0$ a curve , and Euler's angles  $\varphi=\varphi_0,~\psi=\psi_0,~\theta=\theta_0$ \-- initial orientation of the attached triad.
Three Euler's angles form in each point $M$ of a curve internal space anholonomic rotational coordinates, which, as it follows from the equations  (\ref{10})-(\ref{15}),  define the physical dynamics of an oriented  material point.
Passing to the time parameter $t$ in the equations (\ref{10})-(\ref{12}), we will get
\be{10a} v_x(\varphi,\psi,\theta)=\frac{dx}{dt}=v(\cos{\varphi}~\cos{\pf} - \sin{\varphi}~ \sin{\pf}~ \cos\theta),\ee
\be{11b}  v_y(\varphi,\psi,\theta)=\frac{dy}{dt}=v(\sin{\varphi}~\cos{\pf} + \cos{\varphi}~\sin{\pf}~\cos\theta),\ee
\be{12c} v_z(\psi,\theta)=\frac{dz}{dt}=v(\sin{\pf}~\sin\theta),\ee
 where$v=ds/dt$ \--- absolute velocity. Dependence of the components of linear velocity from Euler's angels in these equations allows us to approve, that the system of the equations (\ref{10})-(\ref{15})gives the analytical description of Descartes idea that any physical motion is rotation.
Let's write down a Frenet triad  as
$$e^A_{~\alpha}~,$$
Where holonomic  index  $\alpha$ accepts values $1,~2,~3,$
and index $A$ \--- local anholonomic index  accepts values $1,~2,~3,$ designating numbers of  a triad vectors .

 It will be transformed in group of local three-dimensional rotations
  If on the holonomic  coordinate index $\alpha$   the triad $e^A_{~\alpha}$ has the tensor  law of transformation in group of translations $T(3)$
\be{16}e^A_{~\alpha'}=\frac{\partial x_{\alpha'}}{\partial x_\alpha}e^A_{~\alpha},~~~\alpha=1,2,3,~~~\Vert\frac{\partial x_{\alpha'}}{\partial x_\alpha}\Vert\in T(3),\ee
that on the anholonomic local index  $A$  triad $e^A_{~\alpha}$ it will be transformed in group of local three-dimensional rotations $O(3)$
\be{17}e^{A'}_{~\alpha}=\Lambda^{A'}_{~A}e^{A}_{~\alpha},~~~A=1,~2,~3,~~~\Lambda^{A'}_{~A}\in O(3).\ee
At the description of motion of the Frenet triad  the  group  $T(3)$ and $O(3)$T form Poincare's group with semidirect product of translations and rotations, as rotation of vectors of a triad causes translation of its origin \--- $M$ points of M and  vice versa. This fact substantially distinguishes mechanics of an oriented point from the Newtonian mechanics,  as in the new mechanics translational and rotational energies are not additive.

    \vspace{5mm}

{\raggedright \section{Connection of $\chi_1(s)$ and $\h(s)_2$  with Ricci rotational coefficients and
rotational metric}}

\vspace{2mm}

{\sf Statement 1.}
  Torsion $\chi_1 $ and  $\chi_2$  are independent components
of  Ricci rotation coefficients.

 {\sl Proof.} Let's consider six-dimensional manifold of coordinates
$x_{1},  x_{2},  x_{3}, \varphi_{1}, \varphi_{2}, \varphi_{3}$. It is convenient to present it as a vector
 bundle
 with the base formed by translational coordinates $x_{1},  x_{2},  x_{3}$
(let it be Cartesian coordinates) and fibre, specified at each
point  $x_{\alpha}$  ($\alpha=1,2,3$) by three orthonormalized Frenet's reference
vectors
\begin{equation}
                 \label{18}
{\bf e}_{A},~~~~~~~~~~~~~~~~A=1,2,3,
\end{equation}

where $A$ means number of the reference  vector.

 According to Euler's theorem, an infinitesimal rotations around
the three axes of reference vector (\ref{18})  is equivalent to
one rotation with angle ${\bf d}\bfh$ around a definite axis
passing through the origin of the axis $O$ (see (fig.\ref{r2}b) .  It is possible to
define the infinitesimal rotation as
$${\bf d}\bfh=d\h{\bf e}_\h,$$
where vector ${\bf e}_{\h}$  is directed along instantaneous rotation axis
of reference system. This direction is selected so that,
 if one looks from the end of the vector ${\bf e}_{\h}$ e  at a fixed point
$O$, then the rotation is made counter-clockwise (right-hand reference system).

Let's note, that the vector $\bfh$ does not exist, as turn on a finite angle is not commutative. Therefore for an infinitesimal rotation we have entered a designation ${\bf d}\bfh$ instead of $d{\bfh}$. Unlike a polar vector holonomic translational coordinates $d{\bf x}$,  infinitesimal rotation

\be{18a}{\bf d}\bfh={\bf e}_3d\varphi+{\bf e}_\xi d\theta+{\bf e}_{3'}d\psi\ee
is an axial vector.

An infinitesimal rotation of Frenet's reference   vectors ${\bf
e}_{\h}$
 upon rotation ${\bf d}\bfh$ has the form

\begin{equation}
                 \label{19}
d{\bf e}_{A}=[{\bf d}\bfh {\bf e}_{A}].
\end{equation}

 If we divide (\ref{19}) by $ds$, then we shall get
\be{20}
\frac{d{\bf e}_{A}}{ds}=[\frac{{\bf d}\bfh}{ds}{\bf e}_{A}]
=[{\mbox{\bm$\om$}},{\bf e}_{A}],
\ee
where ${\mbox{\bm$\om$}}={\bf d}\bfh /ds$ - three-dimensional angular velocity of
Frenet's triad with respect to the instantaneous axis. Writing
down  the orthogonality conditions  for Frenet's reference vectors in the form

  \begin{eqnarray}
                 \label{21}
a)~~~~~~~e^{A}_{~\al}e^{\al}_{B}={\dl}^{A}_{~B}=
        \left\{
                \begin{array}{ll}
                        1         & A=B\\
                        0         & A\neq B
                \end{array}
        \right.,\\
b)~~~~~~~e^{A}_{~\al}e^{\bt}_{~A}={\dl}_{\al}^{~\bt}=
        \left\{\begin{array}{ll}
                        1         & \al =\bt\\
                        0         & \al\neq\bt
                \end{array}
        \right.,    \nonumber
\end{eqnarray}
$$
A,B...=1,2,3,~~~~~~~~~~~~~~~~~~\al,\dl,\bt=1,2,3,
$$
where  $\al,\dl,\bt...$ \---  holonomic coordinate indices,  and $A,B...$ \---  anholonomic triad local indices;
 it is possible to write down relations (\ref{19}) and (\ref{20}) as follows

\be{22} de^{A}_{~\al}=d\h^{\bt}_{~\al}e^{A}_{~\bt},\ee
\be{23}\frac{de^{A}_{~\al}}{ds}=\frac{d\h^{\bt}_{~\al}}{ds}e^{A}_{~\bt}.\ee
Multiplying (\ref{22}) and (\ref{23}) by $e^{\beta}_{~A}$, we get
\be{22a}d\h^{\bt}_{~\al}=T^{~\bt}_{~\al\g}dx^\g, \ee
\be{24}\frac{de^{A}_{~\al}}{ds}=T^{\bt}_{~\al\g}\frac{dx^\g}{ds}e^{A}_{~\bt}
,\ee where we have defined the designation \be{24}
T^{\al}_{~\bt\g}=e^{\al}_{A}e^{A}_{~\bt,\g}=-e^{A}_{\bt}e^{\al}_{~A,\g}
,~~~T_{\alpha\beta\gamma}=-T_{\beta\alpha\gamma}~~~ ,\g=\frac{\pl}{\pl x^{\g}}. \ee

 The quantities (\ref{24}) were first introduced by G.Ricci
\cite{4} and since then they have been called Ricci rotation
coefficients. Using the orthogonality conditions (\ref{21}) and
the rule of transformation to local indices

$$
T^{A}_{~B\g}=e^{A}_{~\al}T^{\al}_{~\bt\g}e^{\bt}_{~B}~,
$$
let's rewrite equations (\ref{24}) in local indices
\be{25}
\frac{de^{A}_{~\al}}{ds}=T^{A}~_{B\g}\frac{dx^\g}{ds} e^{B}_{~\al}~.
\ee

Let's chose vectors $e^{(1)}{}_\al,   e^{(2)}{}_\al$ and
$e^{(3)}{}_\al$ so, that they coincide with Frenet's vectors, and
thus the vector $e^{(1)}{}_\al=dx_\alpha/ds=t_\alpha$ satisfies the condition $t_\alpha t^\alpha=1$.
Then the equations (\ref{23}) become the  well-known Frenet's
equations (\ref{3}-{5}), in which
\be{26}\kappa=\chi_1(s)=T^{(1)}_{(2)\g}\frac{dx^{\g}}{ds}~,~~~
\chi=\h_2(s)=T^{(2)}_{(3)\g}\frac{dx^{\g}}{ds}~.\ee
While deducing
(\ref{3}-{5}) from (\ref{23}), we used the following relations
\be{27} \frac{dx^{\g}}{ds} = e^{\g}{}_{(1)},~~~e^{\g}_{(1)}e_{\g}^{(1)}=1~.\ee

From the relations (\ref{24}) it is clear, that in Frenet's
equations curvature and torsion are expressed through components
of Ricci rotation coefficients (\ref{22}), that proves the
Statement 1.

 The Ricci rotation coefficients are the part of the connection of absolute
parallelism geometry \cite{5} and have an anti-symmetry on the two lower
indices

$$
T^\al{}_{[\bt\g]} = -\Omega^{..\al}_{\bt\g},
$$
\be{28}
\Omega^{..\al}_{\bt\g} =  -\frac{1}{2}e^{\al}_{A}(e^{A}_{~\bt,\g} - e^{A}_{~\g,\bt}),
\ee
which it is possible to call {\it Ricci torsion}. Let's note, ones more, that the curvature $\kappa$
and torsion $\chi$ of Frenet's curve would be more correctly called  the first and second torsion,
as they are both expressed through components of Ricci torsion (\ref{28}).

From (\ref{27}) we can find $ds=e_{\alpha}{}^{(1)}dx^\alpha$ and
\be{29}ds^2= e_{\alpha}{}^{(1)}dx^\alpha e^{\alpha}{}_{(1)}dx_\alpha=dx^\alpha dx_\alpha=dx^2+dy^2+dz^2.\ee
This translational metrics is set on group $T(3)$ of translational coordinates and defines geometry of the 3D euclidian space, in which the curve is embedded. Besides as follows from (\ref{18a}) and (\ref{25}), on the group of rotational coordinates $O(3)$  the rotational metrics is define.
$${\bf d}\bfh^2=d\varphi^2+d\psi^2+d\theta^2=d\chi^\alpha_{~\beta}d\chi^\beta_{~\alpha}=T^\alpha_{~\beta \gamma}T^\beta_{~\alpha \delta}dx^\gamma dx^\delta=d\tau^2,$$
This metrics addresses in zero if the first and second torsions (\ref{26} address to zero.

            \vspace{5mm}
{\raggedright \section{4D oriented point and absolute parallelism geometry }}
\vspace{5mm}

A 3D oriented  material point is mathematical representation of an arbitrary accelerated three-dimensional system of reference. Motion of such system of reference is described by six equations as it has six degrees of freedom. It would be possible to put and solve the problem on what geometry possess the space of events of an arbitrary accelerated 3D systems of reference. However, we consider as more important  question \ --- what geometry possess  space of events of an arbitrary accelerated 4D systems of reference or, that is the same, what the space of events form  the relative coordinates of the 4D oriented material points?  It is in advance possible to tell, and it is obvious, that 4D an arbitrary  accelerated system of reference has 10 degrees of freedom, therefore, for the description of its motion, it is necessary to use 10 coordinates. Leaning on experience, which we have received at the description of the dynamics of an arbitrary accelerated 3D system, we shall consider 10D coordinate space in which 4 translational  coordinates $x_0=ct,~x_1=x,~x_2=y,x_3=z$  describe motion of the origin $O$ 4D oriented point and  6 angular coordinates
$\varphi_1=\varphi,~\varphi_2=\psi,~\varphi_3=\theta,~\varphi_4=\vartheta_x,~\varphi_5=\vartheta_y,~\varphi_6=\vartheta_z$ describe change of its orientation.

~~~Consider a four-dimensional differentiable manifold of 4D oriented points with translational coordinates $x^i$
($i = 0,1,2,3~~~a = 0,1,2,3$). Whit each point of the manifold we connect four vectors
$e^a_{~i}$    ($i  =  0,1,2,3$) and four covectors $e^j_{~b}$
with the orthogonality conditions
\be{30}
e^a_{~i} e^j_{~a} =\delta_i^j, \quad e^a_{~i} e^i_{~b} = \delta^a_b.
\end{equation}
Anholonomic tetrad  $e^a_{~i}$ defines the   metric tensor of the space
        \begin{equation} \label{31}
g_{ik} = \eta_{ab}e^a_{~i} e^b_k,
\eta_{ab} = \eta^{ab} = {\rm diag} (1\,-1\,-1\,-1)
        \end{equation}
and the translational Riemannian metric
\be{32}
ds^2 = \eta_{ab}e^a_{~i} e^b_kdx^i dx^k=g_{ik}dx^i dx^k. 
\end{equation}
Using the tensor (\ref{31}), we can construct
the Christoffel symbols
\be{31}
\Gamma^i_{jk} = \frac {1}{2}g^{im}(g_{jm,k} + g_{km,j} - g_{jk,m}).
\end{equation}
that transform following a nontensor law of transformation
\be{31}
\Gamma^{k'}_{j'i'} = \frac {\partial^2 x^k}{\partial x^{i'} \partial
x^{j'}} \frac {\partial x^{k'}}{\partial x^k} + \frac {\partial
x^i}{\partial x^{i'}} \frac {\partial x^j}{\partial x^{j'}} \frac
{\partial x^{k'}}{\partial x^k} \Gamma^k_{ji} 
\end{equation}
with respect to  the coordinate transformations
$$
dx^{i'} = \frac{\partial x^{i'}}{\partial x^{k}} dx^k,~~~\Vert\frac{\partial x^{i'}}{\partial x^{k}}\Vert\in T(4).
$$
were $T(4)$ \--- group of 4D translations.
Now the Ricci rotation coefficients (\ref{24})
 can be represented in the form
\be{33}
T^i_{jk} = e^i_{~a} \nabla_k e^a_{~j},~~~T^i_{jk} = - e^a_{~j} \nabla_k e^i_{~a},~~~T_{ijk}=-T_{jik},\ee
where $\nabla_k$ is a covariant derivative with respect to the
Christoffel $\Gamma^i_{jk}$ symbols.
The rotational metric in the new space can be written as
\be{34} d\tau^2=d\chi^a_{~b}d\chi^b_{~a}=T^a_{~bn}T^b_{~am}dx^kdx^m~,\ee
$$i,~j~,k...=0,1,2,3,~~~a,~b,~,c...=0,1,2,3.$$

Let we have an arbitrary curve in four-dimensional Riemannian
space with translational coordinates $x^i, (i=0,1,2,3)$. Then the
curve is defined by three scalar invariants  ${\h}_1,~{\h}_2~
{\h}_3$, and in our case the four-dimensional Frenet's equations  have
 \be{35}
\frac{De^{(0)}_{~k}}{ds}=\h_1e^{(1)}_{~k},\ee \be{54}
\frac{De^{(1)}_{~k}}{ds}=\h_1e^{(0)}_{~k}+\h_2e^{(2)}_{~k},\ee
\be{36}
\frac{De^{(2)}_{~k}}{ds}=-\h_2e^{(1)}_{~k}+\h_3e^{(3)}_{~k},\ee
\be{37} \frac{De^{(3)}_{~k}}{ds}=-\h_3e^{(2)}_{~k}.\ee
 Here vectors $e^{(0)}_{~k}, e^{(1)}_{~k}, e^{(2)}_{~k}$~and~ $e^{(3)}_{~k}$  form a tetrad, and
  $D$ is the absolute differential with respect to the
four-dimensional Christoffel symbols (\ref{31}).

\vspace{2mm} {\sf Statement 2.} Any curve of Riemannian space can
be considered as the geodesics  of space
of absolute parallelism \cite{5}, with equations of the form
\be{38}
\frac{d^2x^{i}}{ds^2}=-\G^{i}_{~jk}\frac{dx^j}{ds}\frac{dx^k}{ds}-
T^{i}_{~jk}\frac{dx^j}{ds}\frac{dx^k}{ds}.\ee

{\sl Proof.} Connection of absolute parallelism is defined as
\cite{3}
 \be{39} \Delta^{i}_{~jk}= \G^{i}_{~jk}+T^{i}_{~jk} =
e^{i}_{a}e^{a}_{~j,k}=- e^{a}_{~j}e^{i}_{~a,k}. \ee
These
relations can be rewritten as follows
\be{40} T^{i}_{~jk} =
e^{i}_{~a}\nb_k e^{a}_{~j}=-e^{a}_{~j}\nb_k e^{i}_{~a}, \ee where
$\nb_k$   - covariant derivative with respect to Christoffel
symbols. Multiplying equality (\ref{40}) on $e^a_{~i}~
(e^j_{~a}) $ and using the orthogonality conditions (\ref{30})
let's present (\ref{40}) as follows
\be{41} a)~~\nb_k
e^{a}_{~j}=T^{a}_{~bk}e^b_{~j}~~~or~~~ b)~~\nb_k
e^{i}_{~a}=-T^{i}_{~jk}e^j_{~a}. \ee Multiplying (41a) and (41b)
on $dx^{k}/{ds}$, we shall obtain
\be{42} \frac{De^{a}_{~j}}{ds}=
T^{a}_{~bk}e^b_{~j}\frac{dx^k}{ds}.\ee

\be{43}
\frac{De^{i}_{~a}}{ds}= - T^{i}_{~jk}e^j_{~a}\frac{dx^k}{ds}.\ee

Uncovering in equations (\ref{43}) the absolute differential and supposing
in them $ e^{i}_{~(0)}=dx^i/ds$, we shall obtain geodesics equations (\ref{38}).

 Changing in equations (\ref{42}) indices on which there is a contraction,
we find
$$\frac{De^{a}_{~k}}{ds^2}= T^{a}_{~bj}e^b_{~k}\frac{dx^j}{ds}.$$
 Choosing in these equations the Frenet's tetrad and writing down
them component by component, we have
\be{44}
\frac{De^{(0)}_{~k}}{ds^2}= T^{(0)}_{~(1)j}e^{(1)}_{~k}\frac{dx^j}{ds}, \ee
 \be{45}
\frac{De^{(1)}_{~k}}{ds^2}= T^{(1)}_{~(0)j}e^{(0)}_{~k}\frac{dx^j}{ds}
+ T^{(1)}_{~(2)j}e^{(2)}_{~k}\frac{dx^j}{ds}, \ee
 \be{46}
\frac{De^{(2)}_{~k}}{ds^2}= T^{(2)}_{~(1)j}e^{(1)}_{~k}\frac{dx^j}{ds}
+ T^{(2)}_{~(3)j}e^{(3)}_{~k}\frac{dx^j}{ds}, \ee
 \be{47}
\frac{De^{(3)}_{~k}}{ds^2}= T^{(3)}_{~(2)j}e^{(2)}_{~k}\frac{dx^j}{ds}. \ee
 Comparing equations (\ref{53})-(\ref{56}) with equations (\ref{64})-(\ref{67}),
we shall obtain
$$
\h_1=T^{(0)}_{~(1)j}\frac{dx^j}{ds},~~~\h_2= T^{(1)}_{~(2)j}\frac{dx^j}{ds},~~~
\h_3=T^{(2)}_{~(3)j}\frac{dx^j}{ds}.
$$
Since the quantities $T^{i}_{~kj}$ are defined through Ricci
torsion (see (\ref{40})), then,  as it follows from relations
obtained above, {\it is possible to  geometrize  any curves of
Riemannian space, using Ricci torsion}.

The common symmetries os space of events of 4D oriented point are determined as:

a)by transformation of the four holonomic translation coordinates $x_i$ , describing the motion of the origin of an arbitrary  accelerated 4D frame
\be{48}e^a_{~i'}=\frac{\partial x_{i'}}{\partial x_i}e^a_{~i},~~~i=0,1,2,3,~~~\Vert\frac{\partial x_{i'}}{\partial x_i}\Vert\in T(4),\ee
where $T(4)$ is a local group of 4D translations;

b) by transformation of the six anholonomic rotational coordinates $\bfh_{ab}=-\bfh_{ba}$, describing rotation of 4D oriented point (or an arbitrary  accelerated 4D frame)
\be{49}e^{a'}_{~i}=\Lambda^{a'}_{~a}e^{a}_{~i},~~~a=0,1,~2,~3,~~~\Lambda^{a'}_{~a}\in O(1.3),\ee
where $O(1.3)$  is a local Lorenz group of 4D rotations. Term "local group" means, that the parameters of the group depends on the point of the curve.

The matrix $\Lambda^{a'}_{~a}$ can be represented as

$$\Lambda^{a'}_{~a}=R^{a'}_{~b}L^b_{~a}$$
where
\be{50}
 R^{a'}_{~b} = \left( \ba{cccc}
1 &                 0 &                 0 &                 0 \\
0 & \cos \varphi_{xx} & \cos \varphi_{xy} & \cos \varphi_{xz} \\
0 & \cos \varphi_{yx} & \cos \varphi_{yy} & \cos \varphi_{yz} \\
0 & \cos \varphi_{zx} & \cos\varphi_{zy} & \cos \varphi_{zz} \ea
\right) , \ee
is the matrix of the spatial rotations and
\be{51}
 L^b_{~a} = \left( \ba{cccc}
\g &    -\bt_x \g  &   -\bt_y \g      &  -\bt_z \g   \\
-\bt_x \g & 1+\frac{(\g-1)  \bt^2_x}{\bt^2}  &
        \frac{(\g-1) \bt_x\bt_y}{\bt^2} & \frac{(\g-1) \bt_x\bt_z}{\bt^2} \\
-\bt_y \g & \frac{(\g-1)  \bt_x\bt_y}{\bt^2} &
      1+\frac{(\g-1) \bt^2_y}{\bt^2} & \frac{(\g-1) \bt_y\bt_z}{\bt^2} \\
-\bt_z \g & \frac{(\g-1)  \bt_x\bt_z}{\bt^2} &
      \frac{(\g-1) \bt_y\bt_z}{\bt^2} & 1+\frac{(\g-1) \bt^2_z}{\bt^2}
\ea  \right) , \ee
\--- is the matrix, which  describes rotation in space-time planes.
Here
 $$ \g=\frac 1{\sqrt{1-\bt^2}}, ~~~
\bt^2=\bt^2_x+\bt^2_y+\bt^2_z $$
\--- is relativistic factor, in which
3D velocity $v_\alpha=dx_\alpha/dt$ of 4D frame connects whit $\beta_\alpha$ and space-time angle $\vartheta_\alpha$ as
 \be{52} \frac{v_{\al}}{c}=\bt_{\al}= th\vartheta_\alpha,\ee
 were $c$ - velocity of light.

  \vspace{5mm}
{\raggedright \section{Generalization of the Einstein's mechanics }}

\vspace{5mm}

Einstein's General Relativity  assumes the description of laws of physics in an arbitrary  accelerated 4D frames. As we have shown, an arbitrary  accelerated 4D frame has 10 digress of freedom an describes by 10 equations of motions: four equations of motions of the origin of 4D frame  (\ref{38}) and  six rotational equations of motion (\ref{43}). Einstein used only four equations
 \be{53}
\frac{d^2x^{i}}{ds^2}=-\G^{i}_{~jk}\frac{dx^j}{ds}\frac{dx^k}{ds}.\ee

\vspace{5mm}
{\raggedright \subsection{Generalization of the equations of motion of accelerated 4D frame   }}

\vspace{5mm}
 The equations of motion of the origin of 4D  oriented point (or  an arbitrary   accelerated 4D frame )
   coincide with the equations of the
geodesics of the space of absolute parallelism
\be{54}
\frac{d^{2}x^i}{ds^2}+\G^{i}_{~jk}\frac{dx^{j}}{ds}\frac{dx^{k}}{ds}+
T^{i}_{~jk}\frac{dx^{j}}{ds}\frac{dx^{k}}{ds}=0, \ee which differ
from the equations of motion  in Einstein's theory of gravitation (\ref{53})
by  the additional term $$T^{i}_{~jk}\frac{dx^{j}}{ds}
   \frac{dx^{k}}{ds}.$$

The  name of the quantities
$$T^{i}_{~\,jk}=e^{i}_{~a}\nb_{k}e^{a}_{~j}$$ \---the Ricci
rotation coefficients suggests that they describe rotation. It
follows from   our analysis, that the quantities $T^{i}_{~jk}$
describe the change in the orientation of the tetrad vectors
$e^{a}_{~j}$ when the origin of tetrad shifts by an infinitesimal distance
$dx^{i}.$

Einstein interpreted symbols $\Gamma^{i}_{~jk}$ in his equations  (\ref{53}) as intensity of a gravitational field.  The object $\Gamma^{i}_{~jk}$ get  transformed relative to the transformations in  $T(4)$ group as nontensor, whit respect formula (\ref{31}). So, using normal coordinates, we can make $\Gamma^{i}_{~jk}$ equal to zero.  The Ricci rotational coefficients under transformation
of translation coordinates in $T(4)$ transform as tensor
\be{55}
T^{k'}_{j'i'} =  \frac {\partial
x^i}{\partial x^{i'}} \frac {\partial x^j}{\partial x^{j'}} \frac
{\partial x^{k'}}{\partial x^k} T^k_{ji} 
\end{equation}
Writing down the equations (\ref{54}) in normal coordinates, we have
\be{56}
\frac{d^{2}x^{i}}{ds^{2}}+
T^{i}_{~jk}\frac{dx^{j}}{ds}\frac{dx^{k}}{ds}=0, \ee

Using the Ricci rotation
coefficients we can form the 4D angular velocity of
 the tetrad vector \be{57}
\Om^{i}_{~\,j}=T^{i}_{~\,jk}\frac{dx^{k}}{ds} \ee with the
symmetry properties
\be{58} \Om_{ij}=-\Om_{ji}.\ee

Suppose now that the tetrad vectors coincide with the vectors of a
4D arbitrarily accelerated reference frame, then, by
(\ref{57}), the rotation of the reference frame is fully
determined by the torsion field $T^{i}_{~jk}$. Since the field
$T^{i}_{~jk}$ transforms following a tensor law relative to the
coordinates trans\-for\-ma\-tions $x_i$, the rotation of reference
frames relative to the coordinate trans\-for\-ma\-tions
 is absolute. The nontensor
transformation law of $T^{i}_{~jk}$ is valid for
trans\-for\-ma\-tions in the angular coordinates $\varphi_{1},~
\varphi_{2},~
\varphi_{3},~ \vartheta_{1},~
\vartheta_{2},~ \vartheta_{3}$, therefore rotation is only relative for the
group of rotations $O(1.3)$ \cite{5}.

Let us now write the nonrelativistic equations of motion of a mass
$m$ under inertia forces alone, assuming that at a given moment of
time it passes through the origin of an accelerated system

\be{58} \frac{d}{dt}(m{\bf v})=m(-{\bf W}+2[{\bf v}\bfom]), \ee
where ${-m\bf W}$\--- force of inertia, arising at forward acceleration and
$2m[{\bf v}\bfom]$ \--- Coriolis force of inertia.

These equations can be written in the form \be{59}
\frac{d}{dt}(mv_{\al})=m(-W_{\al
o}+2\om_{\al\bt}\frac{dx^{\bt}}{dt}), ~~~\al,\bt=1,2,3, \ee where
${\bf W} = ( W_{1}, W_{2} , W_{3})= (W_{10}, W_{20}, W_{30}),
 \bfom= (\om_{1} ,\om_{2} ,\om_{3})$,
\be{60} \om_{\al\bt}=-\om_{\bt\al}=-\left(
        \begin{array}{ccc}
0    & -\om_{3}   & \om_{2} \\
\om_{3}   & 0      & -\om_{1} \\
-\om_{2}   & \om_{1}  & 0
         \end{array}
 \right).
\ee On the other hand, equations (\ref{56}),  if we take into
account (\ref{57}), can be represented as \be{61}
\frac{d^{2}x^{i}}{ds^{2}}+\Om^{i}_{~\,j}\frac{dx^{j}}{ds}=0. \ee

Multiplying these equations by mass $m$, we will write the
nonrelativistic three-dimensional part of these equations in the
form \be{62} m\frac{du_{\al}}{ds_{0}}=-m\Om_{\al
0}\frac{dx^{0}}{ds_{0}}-2m\Om_{\al\bt} \frac{dx^{\bt}}{ds_{0}}.
\ee

Since in a nonrelativistic approximation
$$
 ds_{0}=cdt, u_{\al}=\frac{v_{\al}}{c}
 $$
and $dx_{0}=cdt$, then the equations (\ref{62}) become \be{63}
m\frac{dv_{\al}}{dt}=-mc^{2}\Om_{\al
0}-2mc^{2}\Om_{\al\bt}\frac{1}{c} \frac{dx^{\bt}}{dt}. \ee
Comparing (\ref{63}) with (\ref{59}) gives \bean
\Om_{10}=\frac{W_{1}}{c^{2}},~~~\Om_{20}=\frac{W_{2}}{c^{2}}, \\
\Om_{30}=\frac{W_{3}}{c^{2}},~~~\Om_{12}=-\frac{\om_{3}}{c}, \\
\Om_{13}=\frac{\om_{2}}{c},~~~\Om_{23}=-\frac{\om_{1}}{c}. \eean

Consequently, the matrix of the 4D angular velocity
of rotation of an arbitrarily accelerated reference frame (matrix
of the 4D "classical spin") has the form
    \begin{widetext}
\be{64}
\Om_{ij}=\frac{1}{c^{2}} \left( \ba{cccc}
O    & -W_1    & -W_2    & -W_2  \\
W_{1}  & 0     & -c\om_{3}     & c\om_{2}  \\
W_{2}  & c\om_{3}   & 0   & -c\om_{1}  \\
W_{3}  & -c\om_{2}  & c\om_{1}   & 0 \ea \right)=\frac{1}{c} \left( \ba{cccc}
O    & -\Theta_1    & -\Theta_2    & -\Theta_2  \\
\Theta_{1}  & 0     & -\om_{3}     & \om_{2}  \\
\Theta_{2}  & \om_{3}   & 0   & -\om_{1}  \\
\Theta_{3}  & -\om_{2}  & \om_{1}   & 0 \ea \right), \ee  \end{widetext}
were $\omega_\alpha=d\varphi_\alpha/dt,~~~\alpha=1,2,3$ \--- spatial angular velocity and
$\Theta_\alpha=d\vartheta_\alpha/dt,~~~\alpha=1,2,3$\--- angular velocity in the space-time planes.
So,  in nonrelativistic approximation 3D acceleration of 4D frame origin

$$W_\alpha=c\Theta_\alpha=c\frac{d\vartheta_\alpha}{dt},~~~\alpha=1,2,3$$
looks like rotation in the space-time planes.
It is seen from the matrix that the 4D rotation of a
 frame caused by the inertial fields $T^{i}_{~jk}$ is
associated with the torsion  \be{65}
\Delta^i_{~\,[jk]} =T^{i}_{~\,[jk]}=- \Om_{jk}^{\,.\,.\,i}=-e^{i}_{~\,a}e^{a}_{~[k,j]}=-
\frac{1}{2}e^{i}_{~\,a}(e^{a}_{~k,j}-e^{a}_{~\,j,k})\ee  of a space of absolute parallelism,
 since
 \be{66}
T^i{}_{jk} = -\Omega^{..i}_{jk} + g^{im} (g_{js}\Omega^{..s}_{mk} +
g_{ks} \Omega^{..s}_{mj}).
\ee

 Fields determined by the rotation of
space came to be known as torsion fields. Accordingly, the torsion
field $T^{i}_{~jk}$  represents the inertial  field engendered by
the torsion of a space of absolute parallelism \cite{ein}. A. Einstein used the torsion
of absolute parallelism  determining torsion as
$$\Lambda^i_{~jk}=\frac{1}{2}(\Delta^i_{~jk}-\Delta^i_{~kj}),$$
where $$\Delta^i_{~jk}=e^i_a
e^a_{j,k},~~~{,k} = \frac {\partial}{\partial
x_{k}},~~~i,j,k...=0,1,2,3,$$
\---connection of absolute parallelism,
$$\Lambda^i_{~jk}=-\Om_{jk}^{\,.\,.\,i}$$ \--- anholonomity object
in J. Schouten definition. In the same article  A.  Einstein has
specified, that when torsion  $\Lambda^i_{~jk}$  (
anholonomity object ) is equal to zero the space becomes Minkovski
space.

\vspace{5mm}
{\raggedright \subsection{Generalization of the Einstein' vacuum equations}}
\vspace{5mm}
An empty, but curved space in Einstein's theory obeys the
equations
\begin{equation}
                         \label{67}
R_{ik}=0,
\end{equation}
whose Schwarz\-schild's solution  is supported by experiment (the shift
of Mercury's perihelion, the deviation of a light ray in the Solar gravitational
field, the delay of radiosignals in a gravitational field, etc.).

Note that Einstein's vacuum equations {\it do not contain any physical
constants}. They are purely field nonlinear equations, and Einstein held that
a correct generalization of these equations would lead us to equations of the
unified field theory. He wrote \cite{6}: "I believe, further, that the equations
of gravitation for empty space are the only rationally justified case of field
theory that can claim to be rigorous (considering nonlinear terms as well).
This all leads to an attempt to generalize the gravitation theory for empty
space."

Einstein believed that one of the main problems in
unified field theory is the one of the geometrization of the
energy-momentum tensor of matter on the right-hand side of his
equations (\ref{1}). This problem can be solved using the concept of 4D oriented point
and the space of events with  the geometry of
absolute parallelism and Cartan's structural equations in this
geometry \cite{5}:

Displacement of the origin and changing of the orientation of 4D oriented point
can be presented by differentials
\be{68}
dx^i = e^a e^i_{~a},
\end{equation}
\be{69}
de^i_{~b} = \Delta^a_{~b} e^i_{~a},
\end{equation}
where
\be{553}
e^a = e^a_i dx^i,
\end{equation}
\be{70}
\Delta^a_{~b} = e^a_{~i} de^i_{~b} = \Delta^a_{~bk} dx^k 
\end{equation}
are differential 1-forms of tetrad $e^a_{~i}$ and connection of absolute
parallelism $\Delta^a_{~bk}$. Differentiating the relationships  (\ref{68}),
(\ref{69}) externally, we have, respectively,
\be{71}
d(dx^i) = (de^a - e^c \wedge \Delta^a_{~c})e^i_{~a} =- S^a e^i_{~a}, 
\end{equation}
\be{72}
d(de^i_{~a}) = (d\Delta^b_{~a} -\Delta^c_{~a} \wedge \Delta^b_{~c})e^i_{~b} =- S^b_{~a}
e^i_{~b}.
\end{equation}

Here $S^a$ denotes the 2-form of Cartanian torsion, and $S^b_{~a}$
\--- the 2-form of the curvature tensor. The sign $\wedge{}$ signifies external
product, e.g,
\be{73}
e^a \wedge e^b = e^a e^b - e^b e^a.
\end{equation}

By definition, a space has a geometry of absolute parallelism, if the 2-form
of Cartanian torsion $S^a$ and the 2-form of the Riemann-Christoffel curvature
$S^b_{~a}$ of this space vanish
\be{74}
S^a = 0, 
\end{equation}
\be{75}
S^b_{~a} = 0. 
\end{equation}

At the same time, these equalities are the integration conditions for the
differentials (\ref{74}) and (\ref{75}).

Equations
\be{76}
de^a - e^c \wedge \Delta^a_{~c} = - S^a, 
\end{equation}
\be{77}
d\Delta^b_{~a} -\Delta^c_{~a} \wedge \Delta^b_{~c} = - S^b_{~a}, 
\end{equation}
which follow from (\ref{71}) and (\ref{72}), are Cartan's structural
equations for an appropriate geometry. For the geometry of absolute parallelism
hold the conditions (\ref{74}) and (\ref{75}), therefore Cartan's structural equations
for $A_4$ geometry have the form
\be{78}
de^a - e^c \wedge \Delta^a_{~c} = 0,
\end{equation}
\be{79}
d \Delta^b_{~a} - \Delta^c_{~a} \wedge \Delta^b_{~c} = 0. 
\end{equation}
Considering (\ref{39}), we will represent 1-form $\Delta^a_{~b}$ as the sum
\be{80}
\Delta^a_{~b} = \Gamma^a_{~b} + T^a_{~b}. 
\end{equation}

Substituting this relationship into (\ref{78}) and noting that
\[ e^c \wedge \Delta^a_{~c} = e^c \wedge T^a_{~c}, \]
we get the first of Cartan's structural equations for space of events of the 4D oriented points
$$
de^a - e^c \wedge T^a_{~c} = 0.
$$
In matrix form these equations will look like
$$
\nabla_{[k} e^a_{~m]} - e^b_{~[k} T^a_{~|b|m]} = 0.\eqno (A)
$$

Substituting (\ref{80}) into (\ref{79}) gives the second of Cartan's
equations for the space.
$$
R^a_b + dT^a_{~b} - T^c_{~b} \wedge T^a_{~c} = 0,
$$
or, in matrix form
$$
R^a_{~bkm} + 2\nabla_{[k} T^a_{|b|m]} + 2T^a_{~c[k} T^c_{~|b|m]} = 0. \eqno (B)
$$

 In the coordinate indexes the equations $(B),$ written as
\be{81}
R^i_{~jkm} + 2\nabla_{[k}T^i_{|j|m]} + 2T^i_{s[k} T^s_{|j|m]} = 0. 
\end{equation}

Forming, using (\ref{81}), the Einstein tensor
$$ G_{jm} = R_{jm} - \frac {1}{2} g_{jm}R, $$
we obtain the 10 equations
\be{82}
R_{jm} - \frac {1}{2} g_{jm} R = \nu T_{jm}, 
\end{equation}
which are similar to Einstein's equations, but with the geo\-met\-rized
right-hand side defined as
\begin{eqnarray} \lb{83}
 T_{jm}=-\frac{2}{\n}\{(\nabla_{[i}T^{i}_{~|j|m]}+T^{i}_{~s[i}T^{s}_{~|j|m]}) - \nonumber\\
-\frac{1}{2}g_{jm}g^{pn}(\nabla_{[i}T^{i}_{~|p|n]}+T^{i}_{~s[i}T^{s}_{~|p|n]})\}
\end{eqnarray}
Let us now decompose the Riemann tensor $R_{ijkm}$ into irreducible parts
\be{84}
R_{ijkm} = C_{ijkm}+g_{i[k}R_{m]j}+g_{j[k}R_{m]i}+\frac {1}{3}
Rg_{i[m}g_{k]j},
\end{equation}
where $C_{ijkm}$ is the Weyl tensor; the second and third terms are the
traceless part of the Ricci tensor $R_{jm}$ and $R$ is its trace.

Using the equations (\ref{82}), written as
\be{85}
R_{jm} = \nu \left( T_{jm} - \frac {1}{2} g_{jm} T \right), 
\end{equation}
we will rewrite the relationship (\ref{84}) as
\be{86}
R_{ijkm} = C_{ijkm} + 2\nu g_{[k(i} T_{j)m]} - \frac {1}{3} \nu
Tg_{i[m}g_{k]j}, 
\end{equation}
where $T$ is the tensor trace (\ref{83}).

Now we introduce the tensor current
\be{87}
J_{ijkm}= 2g_{[k(i} T_{j)m]} - \frac {1}{3} Tg_{i[m} g_{k]j} 
\end{equation}
and represent the tensor (\ref{86}) as the sum
\be{88}
R_{ijkm} = C_{ijkm} + \nu J_{ijkm}. 
\end{equation}

Substituting this relationship into the equations (\ref{81}), we will arrive
at
\be{89}
C_{ijkm} + 2\nabla_{[k}T_{|ij|m]} + 2T_{is[k}T^s_{|j|m]} = - \nu
J_{ijkm}. 
\end{equation}

Equations (\ref{89}) are the Yang-Mills equations  with a geo\-met\-rized
source, which is defined by the relationship (\ref{87}). In equations
(\ref{89}) for the Yang-Mills field we have the Weyl tensor $C_{ijkm}$, and
the potentials of the Yang-Mills field are the Ricci rotation coefficients
$T^i_{jk}$.

Summarizing the geo\-met\-rized Einstein equations (\ref{82}) and the Yang-Mills
equations (\ref{89}), we can represent the structural Cartan equations
$(A)$ and $(B)$ as an extended set of Einstein-Yang-Mills equations

$$\nabla_{[k} e^a_{j]} + T^i_{[kj]} e^a_{~i} = 0,\eqno(A) $$
$$R_{jm} - \frac{1}{2} g_{jm} R = \nu T_{jm},\eqno(B.1)$$
$$C^i_{jkm} + 2\nabla_{[k} T^i_{|j|m]} + 2T^i_{s[k} T^s_{|j|m]} = - \nu J^i_{~jkm}, \eqno(B.2)$$
in which the geo\-met\-rized sources $T_{jm}$   and $J_{ijkm}$ are given by
(\ref{83}) and (\ref{87}).

For the case of Einstein's vacuum the equations  are much simpler

$$\nabla_{[k} e^a_{j]} + T^i_{[kj]} e^a_{~i} = 0, \eqno(i)$$
$$R_{jm} = 0, \eqno (ii)$$
$$C^i_{~jkm} + 2\nabla_{[k} T^i_{|j|m]} + 2T^i_{s[k} T^s_{|j|m]} = 0.\eqno(iii)$$

Thus, the structural equations of absolute parallelism  geometry,
represent an extended set of Einstein-Yang-Mills equations with the gauge
translations group $T_4$ defined on the base $x^i$ with the structural
equations  $(A),$ and with the gauge rotational group $O(1.3)$, defined in the
fibre $e^i_{~a}$ with the structural equations in the form of the geo\-met\-rized
Einstein-Yang-Mills equations $(B.1)$ and $(B.2).$

It is easy to see, that when torsion $\Om_{jk}^{\,.\,.\,i}$ (and , hence,  torsion field $T^{i}_{~\,jk}$) in the (A) and (B) equations is equal to zero the space of events becomes Minkovski space. The converse proposition, generally, is incorrect. If to put in the equations (A) and (B) Riemannian curvature equal to zero, we shall receive the equations
\be{89a}\nabla_{[k} e^a_{j]} + T^i_{[kj]} e^a_{~i} = 0,\ee
\be{89b}\nabla_{[k} T^i_{|j|m]} + T^i_{s[k} T^s_{|j|m]} = 0,\ee
which describe so-called primary torsion fields \cite{5}.

\vspace{5mm}
{\raggedright \section{Correspondence with the equations of Einstein's
theory and deterministic quantum mechanics}}
\vspace{5mm}
The equations (\ref{54}) will be transformed to the equations of motions of Einstein's theory of gravitation when the
inertia force in (\ref{54}) becomes zero
\be{90}
F^i_{I}=mT^{i}_{~\,jk}\frac{dx^{j}}{ds}\frac{dx^{k}}{ds}=0, \ee
or, using (\ref{66}) (for $m\neq 0$)
\be{91}
-\Om_{jk}^{\,.\,.\,i}\frac{dx^{j}}{ds}\frac{dx^{k}}{ds}
+g^{im}(g_{gs}\Om_{mk}^{\,.\,.\,s}+g_{ks}\Om_{mj}^{\,.\,.\,s})
\frac{dx^{j}}{ds}\frac{dx^{k}}{ds}=0. \ee
Since $\Om_{mkj}$ is skew-symmetric in indices $m$ and $k$, then
it follows from (\ref{91}) that in inertial reference frames the
torsion  $\Om_{mkj}$ of the space  is skew-symmetrical in
all the three indices
\be{92}
T_{ijk}=-T_{jik}=-T_{ikg}=-\Om_{ijk}, \ee
but not equal to zero and   coincides with torsion  field  $T^i_{~jk}$. The energy-momentum tensor (\ref{83}) in these case is symmetrical in the
indices $j$ and $m$ to yield
\be{93}
T_{jm}=\frac{1}{\n}(\Om_{sm}^{\,.\,.\,i}\Om_{ji}^{\,.\,.\,s}-
\frac{1}{2}g_{jm}\Om_{s}^{\,.\,ji}\Om_{ji}^{\,.\,.\,s}). \ee
In general case torsion $\Om_{jk}^{\,.\,.\,i}$  has $24$
independent components and it can be represented as the sum of three
irreducible parts as follows
\be{93a}
\Om^{i}_{.jk}=\frac{2}{3}\dl^{i}_{~\,[k}\Om_{j]}+\frac{1}{3}\varepsilon^{n}_{~jks}
\hat{\Om}^{\hat s}+\bar{\Om}^{i}_{.jk},
\ee
where
\be{93b}
\Om^{i}_{.jk}=g^{im}g_{ks}\Om_{mj}^{\,.\,.\,s},
\ee
and  $\Om_{j}$ \--- the vector,  $\hat \Om_{j}$ \--- the pseudovector and  $\bar{\Om}_{{.}jk}^{~i}$ \---the traceless
part of torsion  are given by

\be{93c}
\Om_{j}=\Om^{i}_{.ji},
\ee
\be{93d}
\hat{\Om}_{j}=\frac{1}{2}\varepsilon_{jins}\Om^{ins},
\ee
\be{93e}
\bar{\Om}^{s}_{.js}=0,~~~\bar{\Om}_{ijs}+\bar{\Om}_{jsi}+\bar{\Om}_{sij}=0,
\ee
where $\varepsilon_{ijkm}$ is a fully skew-symmetrical Levi-Civita symbol.

Since in inertial reference frames the torsion $\Om_{ijs}$ is skew-symmetrical
in all the three indices, among the irreducible parts of torsion
 in inertial frames only the pseudovector
(\ref{93d}) is nonzero.

We can define the auxiliary pseudovector $h_{m}$ through the field
(\ref{93d}) as follows
\be{93f}
\Om^{ijk}=\varepsilon^{ijkm}h_{m},~~~\Om_{ijk}=\varepsilon_{ijkm}h^{m}
\ee
and write the tensor (\ref{93}) as
\be{93j}
T_{jm}=\frac{1}{2\n}(h_{j}h_{m}-\frac{1}{2}g_{jm}h^{i}h_{i}).
\ee

If the pseudovector $h_{m}$ is light-like, it can be represented as
\be{93h}
h_{m}=\Phi l_{m},~~~l_{m}l^{m}=0,~~~\Phi=\Phi(x^{i}).
\ee
In this case the matter tensor (\ref{93j}) becomes
\be{93g}
T_{jm}=\frac{1}{\n}\Phi^{2}(x^{i})l_{j}l_{m},
\ee
and the density of matter is given by
\be{93k}
\rho=\frac{1}{\n c^{2}}\Phi^{2}(x^{i}).
\ee

If the pseudovector  $h_{m}$ is time-like, it can conveniently be represented
as
\be{93l}
h_{m}=\varphi(x^{i})u_{m},
\ee
where
\be{93m}
u_{m}u^{m}=1
\ee
and $\varphi(x^{i})$ is a scalar quantity.

Substitution of (\ref{93l}) into the tensor (\ref{93j}) yields the
energy-momentum tensor of the form
\be{93n}
T_{jm}=\frac{1}{\n}\varphi^{2}(u_{j}u_{m}-\frac{1}{2}g_{jm})=-\rho c^2(u_ju_m-\frac{1}{2}g_{jm}),
\ee
were
\be{93o}\rho=-\frac{1}{\nu c^2}\varphi^2(x^i)\ee
density of the  matter.
Tensor (\ref{93g}) looks like an energy-momentum tensor {\it of isotropic
radiation}, and the tensor (\ref{93n}) in its structure looks rather like the
energy-momentum tensor {\it of an ideal liquid}.
Thus, in a post-Einstein's approximation the matter density  is defined through squares of torsion fields $\Phi$ and
 $\varphi$ according to (\ref{93k}) and (\ref{93o}).
 
\vspace{5mm}
{\raggedright \section{New  concept  of mass density and a mass}}
\vspace{5mm}
The mass of a particle in the equations of Newton mechanics has no structure and does not depend on velocity of its movement.
The special theory of a relativity has found out, that the mass depends on the velocity of a particle as
$$m=\frac{m_0}{\sqrt{1-v^2/c^2}},$$
were $m_0$ the rest mass.
On the other hand, in the quantum mechanics the mass of a particle depends on the frequency $\omega$ of the wave function  $\psi$ as
$$m=E/c^2=\hbar\omega/c^2.$$
It was shown before that in the general case the purely field
energy-mo\-men\-tum tensor of matter $T_{jm}$ in the equations of (B.1)
is defined through the inertial fields $T^i_{~jk}$ by formula  (\ref{83}).

Using (\ref{83}), we define the matter density $\rho$   as
\be{117}\rho=T/c^2=(g^{jm}T_{jm})c^{-2}=\frac{2g^{jm}}{\n c^2}(\nb_{[i}T^i_{|j|m]}+T^i_{~s[i}T^s_{|j|m]}),
\ee then  the inertial mass
$M$ of a system  to be calculated through the integral
  \be{118}M=\int(-g)^{1/2}\rho dV\ee
 $$=\frac2{\n c^2}\int(-g)^{1/2}\left\{g^{jm}\left(\nb_{[i}T^i_{|j|m]}+T^i_{~s[i}T^s_{|j|m]}\right)\right\}dV, $$
were $dV$\--- 3D volume.
This relationship shows that the inertial mass in new mechanic
{\it is a measure of the inertial (or torsion) field} that forms the matter density.
According to the formula (\ref{57}),  all fields and forces  of inertia are created by the torsion of the space.
When torsion $\Om_{jk}^{\,.\,.\,i}$ of the space addresses in zero, torsion field $T^i_{~jk}$ also is equal to zero and the mass (\ref{118}) disappears.

        \vspace{5mm}
{\raggedright \subsection{1+3 splitting and possibility the mass control}}
\vspace{5mm}

In the local (tetrad)  indices  the equations (A) and (B) has the form

$$\nabla_{[a}e^k_{~b]} =- T^c_{~[ba]}e^k_{~c},\eqno(A)$$
$$R^a_{~bcd}=-2\nabla_{[c}T^a_{~|b|d]}-2T^a_{~f[c}T^f_{~|b|d]}.\eqno(B)$$
$$a,b,c...=0,1,2,3.$$

Using the 1+3 splitting  (with signature -~+~+~+)of the local  spacetime \cite{12}, we can define
$u_a=dx_a/d\tau$ - the timelike local 4- velocity vector, so that $u_au^a=-1, ~~g_{ab}=u_a u_b-h_{ab},~~ ds^2=g_{ab}dx^adx^b=(u_au_b-h_{ab})dx^adx^b=d\tau^2-dl^2,$ $h_{ab}$  - metric tensor of a 3D  surface, orthogonal to the unit vector $u_a.$

In this case we have from equations (A)
$$\nabla_{b}e^0_{~a} =- T^c_{~ab}e^0_{~c}=\nabla_{b}u_{~a}=- T^c_{~ab}u_{~c},\eqno(A')$$
or
\be{119} T^c_{~ab}=u^c\nabla_b u_a=-W_a u_b u^c +\omega_{ab} u^c+\sigma_{ab} u^c+\frac{1}{3}\Theta h_{ab}u^c,\ee 
where
$$\omega_{ab}=\nabla_ {[b}u_{a]} + W_{[a}u_{b]},\eqno(R)$$ - rotation,
$$\sigma_{ab}=\nabla_ {(b}u_{a)}+ \frac{Du_{(a}}{d\tau}u_{b)}-\frac{1}{3}\Theta h_{ab},\eqno(S)$$ - shear,
$$\Theta=\nabla_a u^a,\eqno(E)$$ -expansion (or contraction) and
$$W_a=u_b\nabla_b u_a=\frac{Du_a}{d\tau}\eqno(W)$$ - the local 4-  acceleration vector, are determined through the irreducible parts (\ref{93c})- (\ref{93e}) of the torsion $\Omega^{..a}_{bc},$
because
$$
T^a{}_{bc} = -\Omega^{..a}_{bc} + \eta^{af} (\eta_{bd}\Omega^{..d}_{fc} +
\eta_{cd} \Omega^{..d}_{fb}).
$$
Substitution (\ref{119}) in the equations $(B)$, gives

$$R^d_{~abc}=2W_a(\omega_{bc}-W_{[b}u_{c]})u^d+
2\nabla_{[c}W_{|a|}u_{b]}u^d-$$
$$-2\nabla_{[c}\omega_{|a|b]}u^d-2\nabla_{[c}\sigma_{|a|b]}u^d-\frac{2}{3}\Theta _{,[c}h_{b]a}u^d
+\frac{2\Theta}{3}[u_a\omega_{bc}-$$
$$-u_aW_{[b}u_{c]}+\omega_{a[c}u_{b]}+\sigma_{a[c}u_{b]}+
\frac{\Theta}{3}h_{a[c}u_{b]}]u^d,\eqno(B^{1+3})$$
 
 Using the equations (B.1) and the energy-momentum tensor
with the structure $T_{ab}=\rho c^2u_au_b$, we have for the matter energy density
$$\rho=\frac{1}{c^2}T_{ab}u^au^b=\frac{1}{\nu c^2}(\nabla_aW^a+2\omega^2-2\sigma^2-\frac{d\Theta}{d\tau}-\frac{1}{3}\Theta^2)$$
and for the mass (M)
$$M=\frac{1}{\nu c^2}\int(-g)^{1/2}(\nabla_aW^a+2\omega^2-2\sigma^2-\frac{d\Theta}{d\tau}-\frac{1}{3}\Theta^2)dV,\eqno(M^{1+3})$$
were
$$\omega^2=\omega_{ab}\omega^{ab},~~\sigma^2=\sigma_{ab}\sigma^{ab}.$$
The mass $(M^{1+3})$ describes the general case for Alcubierre's version of the warp bubble \cite{4}.
From $(M^{1+3})$  follows, that the null energy condition  obeys, when $\nu\geq$ and
$$\nabla_aW^a+2\omega^2- 2\sigma^2-\frac{d\Theta}{d\tau}-\frac{1}{3}\Theta^2\geq0.$$
The practical interest is represented  the constraint equations
$$\nabla_a\omega^a-W_a\omega^a=0$$
connecting acceleration $W_a$ of the center of mass of system
with its internal rotation $2\omega_a=\varepsilon^{abc}\omega_{bc}$.

Let $\sigma=\Theta=0$ equal to zero, then the mass is
$$M(\omega) =\frac{1}{\nu c^2}\int(-g)^{1/2}(\nabla_aW^a+2\omega^2)dV.$$
This equation shows, that we can   operate by mass $M$ , i.e. angular velocity $\omega$ of rotation of masses  which mass $M$ consists of, it will
lead us to the creation of an inertial propulsion system of an essentially new type, that will  move  in space
\cite {13} like Alcubierre's bubble \cite{14} without action on it of an external forces. The elementary scientific model of such  propulsion system - 4D gyroscope  was created in Thailand and the patent for the vacuum torsion propulsion system representing a warp drive was granted.
Now we shall write down Einstein's well-known formula $E=mc^2$ as

\be{1000}E=m(\omega)c^2,\ee
where the mass of object depends on the internal rotation.

    \vspace{3mm}
{\large\bf Conclusion} \vspace{3mm}

The majority fundamental physical theories uses a point manifold and, accordingly, concept of a material point in the basis. Attempts undertaken by theorists to solve such problem as unification  the general theory of a relativity with  quantum theory of a matter till now have not brought success. The basic result of the given work consists in showing an opportunity for decision of this problem, being based on a  manifold  of an oriented points. The basis for the new physics  has been prepared  by outstanding mathematicians \--- G.Frenet, G. Ricci, L.Euler, E.Cartan  and R.Penrose. The basic of the physical ideas have been stated by W.Clifford, E.Mach, in greater degree by A.Einstein, W.Heisenberg and P.Dirac.

 More than 300 years we have been applying Newton's mechanics to explain
non-relativistic mechanical experiments.
Although Newton's mechanics has been generalized three times: by
the special relativity theory, general relativity theory, and
quantum mechanics, there remains a possibility for its further
generalization. The fourth generalization of Newtonian mechanics has become
possible with regards that new mechanics  has been based
upon the following: 1)  Clifford-Einstein  program for geometrization of all physics equations, including classical mechanics,
(Unified Field Theory \cite{2}); 2) Cartan's idea about the connection of the torsion of space with
physical rotation \cite{15}.

Einstein assumed the solution of these  problems in the geometrization of the right hand of its equations.
Generalizing Einstein's vacuum equations, we have introduce structural Cartan equations  geometry of absolute parallelism
as the new vacuum equations.  It has allowed us  not only to find a general view of geometrized energy-momentum tensor, but also to specify  communication  torsion of the space of absolute parallelism with a field of inertia and wave function of quantum mechanics according to Einstein ideas. The mass of any object in the generalized theory
has cleanly field nature and  is defined as a measure of  field of inertia. The rest mass of such object can be operated, using rotation of masses of which the object consists. The first experimental acknowledgement of these theoretical conclusions are already received by us at research of the dynamics so called 4D gyroscope \cite{13}.

{}

\end{document}